\documentclass[12pt]{amsart}
\usepackage{amssymb, amsmath, amsthm}
\theoremstyle{plain}

\theoremstyle{definition}
\newcommand{\eq}{\eqref}

\newcommand{\Ga}{\Gamma}

\newcommand{\ka}{\kappa}
\newcommand{\La}{\Lambda}
\newcommand{\la}{\lambda}

\newcommand{\Om}{\Omega}
\newcommand{\de}{\delta}

\newcommand{\pa}{\partial}

\newcommand{\no}{\nonumber}

\begin{document}

\title[Dispersionless Harry Dym Hierarchy ]
 {Hydrodynamic Reductions of Dispersionless Harry Dym Hierarchy}

\author[Jen-Hsu Chang]{{ Jen-Hsu Chang }\\
  { Department of Computer Science ,\\ Chung-Cheng Institute of Technology,\\
   National Defense University,\\ Taoyuan, Taiwan \\
   E-mail: jhchang@ccit.edu.tw\\
\\
\\
}}
       
\maketitle
\begin{abstract}
We investigate the reductions of dispersionless Harry Dym hierarchy to systems of finitely many partial differential equations. These equations must satisfy the compatibility condition and they are diagonalizable  and semi-Hamiltonian. By imposing a further constraint, the compatibility is reduced to a system of algebraic equations, whose solutions are described.
\end{abstract}
\section{Introduction}
\indent Dispersionless integrable equations arise in various contexts and have  attracted people to investigate it from different point of view, such as
topological field theory(WDVV equation) \cite{Ak,ct2, du1,DZ, kr1}, matrix models \cite{bx, bpx}, conformal maps and interface dynamics\cite{kw,ww,wz}, Einstein-Weyl space\cite{dm,dt}. The hydrodynamic reductions are the most developed method to 
find the exact solutions of the dispersionless integrable equations \cite{bm,Car2, ct1, gt,gt1,KG, ko, kr1}. From the hydrodynamic reductions, one can construct the Riemann's invariants and the corresponding characteristic speeds satisfy the semi-Hamiltonian property or
Tsarev's condition \cite{t}. Hence the generalized hodograph method can be used to find the exact solutions. Also, the solutions of dispersionless integrable equations can be found by nonlinear Beltrami equation \cite{km,km1}, slightly different from the 
generalized hodograph method. \\
\indent Let's recall dispersionless non-standard Lax hierarchy \cite{Li,TT1}. Suppose that  $\la$ is  an algebra of Laurent series of the form
 \[
\la=\{A|A=\sum_{i=-\infty}^Na_ip^i\},
 \]
  with coefficients $a_i$
depending on an infinite set of variables $t_1
 \equiv x, t_2, t_3, \cdots$. We can define a Lie-Bracket associated with $\la$ as follows:
 \[
\{A, B \}=\frac{\pa A}{\pa x}\frac{\pa B}{\pa
p}-\frac{\pa A}{\pa p}\frac{\pa B}{\pa x}, \qquad A,B \in \la
 \]
which can be regarded as the Poisson bracket defined in the
2-dimensional phase space $(x, p)$.
The algebra $\la$ can be decomposed into the Lie sub-algebras as
 \[
\la=\la_{\ge k}\oplus \la_{< k},\qquad (k=0,1,2.)
 \]
 where
\begin{eqnarray} 
\la_{\ge
k}&=&\{A\in \La| A=\sum_{i\ge k}a_ip^i\}\no\\
 \la_{< k}&=&\{A\in\la| A=\sum_{i< k}a_ip^i\}.\no
 \end{eqnarray}
Based on this, the Lax formulation of the  dispersionless integrable hierarchy can be formally defined as 
\[ \frac{\pa \la}{ \pa t_n}= \{(\la^{\frac{n}{N}})_{\ge k}, \la \}. \]
\begin{itemize}
\item For $k=0$, it's called dispersionless KP hierarchy(dKP) \cite{TT1}
\item For $k=1$, it's called dispersionless modified KP(dmKP)  hierarchy \cite{ct, ma} 
\item For $k=2$, it's called dispersionless Harry Dym (dDym) hierarchy \cite{ct,ct1,ma}. It's the purpose of this  paper.
\end{itemize}
We define the dispersionless Harry Dym (dDym)hydrodynamic systems as follows.
The Lax operator of dDym has the form
\[\la(p)=A_{-1}p+A_{0}+A_1 p^{-1}+A_2 p^{-2}+A_3 p^{-3}+\cdots. \]
Then the dDym hydrodynamic system is \cite{ct, ma}
\begin{equation}
\frac{\pa \la}{\pa t_n} =\{\la, \Om_{n}(p)\}, \label{dy}
\end{equation}
where 
\[\Om_{n}(p)=(\la(p)^{n})_{\geq 2}.\]
Here $()_{\geq 2}$ denote the projection of Laurent series onto a linear combination of $\la(p)^n$ with 
$n \geq 2$. From the zero curvature equation ($t_3=t$ and $t_2=y$)
\[ \frac{\pa \Om_{2}(p)}{\pa t}-\frac{\pa \Om_{3}(p)}{\pa y}=\{\Om_{2}(p),\Om_{3}(p)\}, \]
where 
\[\Om_{2}(p)=A_{-1}^2 p^2 \quad  \mbox{and} \quad  \Om_{3}(p)=A_{-1}^3 p^3+3A_{-1}^2 A_0 p^2 , \]
one can get the dispersionless Harry Dym Equation \cite{ct}
\[\frac{\pa A_{-1}}{\pa t}=\frac{3}{4}\frac{1}{A_{-1}}\left[A_{-1}^2\pa_{x}^{-1} 
\left(\frac{A_{-1y}}{A_{-1}^2}\right)\right]_y.\]
Now, considering the $y$-flow, we have
\begin{equation} 
\la_y=\{\la, A_{-1}^2p^2\}=2p A_{-1}^2\la_x-(A_{-1}^2)_x p^2 \la_p, \label{la}
\end{equation}
or 
\begin{eqnarray}
A_{-1y} &=&2A_{-1}^2 A_{0x} \no \\
A_{0y} &=& 2A_{-1}^2 A_{1x}+(A_{-1}^2)_{x}  A_1 \no \\
&\vdots& \no \\
A_{ny}&=& 2A_{-1}^2 A_{n+1, x}+(n+1)(A_{-1}^2)_{x}  A_{n+1}, \label{ha}
\end{eqnarray}
where $n=-1,0,1,2,3 \cdots.$  Comparing it with the Benney moment chain \cite{g,pa, Z}, one calls  \eq{ha} the dDym moment chain. It's the subject of this paper. \\
\indent The paper is organized as follows. In the next section, one considers the reduction problems and obtains finitely many partial differential equations.  In section 3, we prove the semi-Hamiltonian property when using Riemann invariants.
In section 4, by imposing a further constraint, one gets  some particular reductions. In the final section,  we discuss
some problems to be investigated.
\section{The compatibility conditions}
\indent In this section, we consider the hydrodynamic reduction problems following \cite{as,gt,gt1}. For non-hydrodynamic reductions, one refers to \cite{as1}. \\
\indent We assume that the moments $A_i$ are functions of only $N$ independent 
variables $u_i$. If the $A_i$ satisfy \eq{ha}, then it's straightforward to show that the mapping
\[(u_{-1},u_0, u_1,u_2, \cdots, u_{N-2}) \to (A_{-1}, A_0, A_1, A_2, \cdots, A_{N-2})\]
is non-degenerate. Hence without loss of generality we set $u_{-1}=A_{-1}, u_0=A_0, u_1=A_1, \cdots, u_{N-2}=A_{N-2}$. The first $N$-moments are the independent variables, while the higher moments are functions of them, i.e.,
\[A_{k}=A_k(A_{-1}, A_{0}, \cdots, A_{N-2}), \quad k \geq N-1.\]
The equations of motion for $A_{-1}, A_0, \cdots, A_{N-2}$ become
\begin{eqnarray}
\frac{\pa A_j}{\pa y}&=&2A_{-1}^2 A_{j+1,x}+(j+1)(A_{-1}^2)_x A_{j+1}, \quad j \leq N-3 \label{c1} \\
\frac{\pa A_{N-2}}{\pa y}&=&2 A_{-1}^2 A_{N-1,x}+(N-1)(A_{-1}^2)_x A_{N-1} \label{c2} \\
&=&2A_{-1}^2 \frac{\pa A_{N-1}}{\pa A_j}\frac{\pa A_j}{\pa x}+(N-1)(A_{-1}^2)_x A_{N-1}, \no
\end{eqnarray}
while each higher moment $(A_{N-1}, \cdots,)$ must satisfy the overdetermined system ($k \geq N-1 $) using \eq{c1}\eq{c2}
\begin{eqnarray*}
\frac{\pa A_k}{\pa y} &=& \sum_{j=-1}^{N-3}  \frac{\pa A_k}{\pa A_j}[2A_{-1}^2A_{j+1,x}+(j+1)(A_{-1}^2)_x A_{j+1}] \\
&+& \frac{\pa A_k}{\pa A_{N-2}}[2A_{-1}^2 \sum_{j=-1}^{N-2}\frac{\pa A_{N-1}}{\pa A_j}\frac{\pa A_j}{\pa x}+(N-1)(A_{-1}^2)_x A_{N-1}] \\
&=& 2A_{-1}^2[ \sum_{j=-1}^{N-2}\frac{\pa A_{k+1}}{\pa A_j}\frac{\pa A_j}{\pa x}]+(k+1)(A_{-1}^2)_x A_{k+1}. 
\end{eqnarray*}
Comparing the coefficients of $\frac{\pa A_j}{\pa x}(j=-1,0,1,\cdots, N-2)$, one has 
\begin{eqnarray}
\frac{\pa A_{k+1}}{\pa A_{j}}&=& \frac{\pa A_k}{\pa A_{j-1}}+\frac{\pa A_{k}}{\pa A_{N-2}}\frac{\pa A_{N-1}}{\pa A_j},
\quad  0\leq j\leq N-2 \label{j1} \\
A_{-1}\frac{\pa A_{k+1}}{\pa A_{-1}}&=& \sum_{j=-1}^{N-2} \frac{\pa A_k}{\pa A_j}(j+1)A_{j+1}+A_{-1}\frac{\pa A_{k}}{\pa A_{N-2}}\frac{\pa A_{N-1}}{\pa A_{-1}}\\ \label{j2}
&-&(k+1)A_{k+1} \no
\end{eqnarray}
Now, letting $k=N-1$ and defining $r=\log A_{-1},$
 one has 
\begin{eqnarray}
\frac{\pa A_N}{\pa A_{j}}&=& \frac{\pa A_{N-1}}{\pa A_{j-1}}+\frac{\pa A_{N-1}}{\pa A_{N-2}}\frac{\pa A_{N-1}}{\pa A_j},
\quad  0\leq j\leq N-2 \label{n1} \\
\frac{\pa A_{N}}{\pa r}&=& \sum_{j=-1}^{N-2} \frac{\pa A_{N-1}}{\pa A_j}(j+1)A_{j+1}+\frac{\pa A_{N-1}}{\pa A_{N-2}}\frac{\pa A_{N-1}}{\pa r} \label{n2} \\
&-&NA_{N} \no
\end{eqnarray}
The compatibility of \eq{n1} and \eq{n2} gives a system $\Ga $ of $\frac{N(N-1)}{2}$ non-linear second order equation for the single known $A_{N-1}(A_{-1}, A_0, A_1, \cdots, A_{N-2})$. One can show that by induction if $\Ga$ is satisfied then the analogous compatibility for $A_k(k\geq N)$ is also derived. Let's investigate the case $N=2$ in more details.
Then we have ($A_0=s$)
\begin{eqnarray*}
\frac{\pa A_2}{\pa s}&=& A_{1r}\exp (-r)+A_{1s}^2 \\
\frac{\pa A_{2}}{\pa r}&=& A_1 A_{1s}+A_{1s}A_{1r}-2A_2.\\ 
\end{eqnarray*}
Cross-differentiating, one gets the quasi-linear second differential equation
\begin{eqnarray}
&&A_{1rr} \exp(-r)+A_{1s}A_{1rs}-(A_1+A_{1rs})A_{1ss}+A_{1r}\exp(-r)  \label{mo}  \\
&&+(A_{1s})^2=0. \no
\end{eqnarray}
Letting 
\[A_1=a,\quad  A_{1r}=b, \quad  A_{1s}=c,\]
one can express \eq{mo} as the degenerate non-homogeneous hydrodynamic system
\begin{eqnarray}
 \left(\begin{array}{c}
a \\ b\\ c \\ 
\end{array}
\right)_r
&=&\left(\begin{array}{ccc}
0 & 0&0 \\0 & -c \exp r &(a+b)\exp r \\
0&1&0 
\end{array} \right)
\left(\begin{array}{c}
a \\ 
b \\
c \\
\end{array} \right)_s  \label{non} \\
&+ &\left(\begin{array}{c}
b\\-b-c^2 \exp r \\
0
\end{array} \right) \no
\end{eqnarray}
Define the characteristic speeds as 
\begin{eqnarray*}
u&=& \frac{-c \exp r+ \sqrt{c^2 \exp (2r)+4(a+b)\exp r }}{2} \\
v&=& \frac{-c \exp r+ \sqrt{c^2 \exp (2r)-4(a+b)\exp r }}{2}. 
\end{eqnarray*}
A simple calculation yields,
\begin{equation}
 \left(\begin{array}{c}
u \\ v\\ 
\end{array}
\right)_r
=\left(\begin{array}{cc}
v& 0 \\
0&u
\end{array} \right)
\left(\begin{array}{c}
u \\ 
v \\
\end{array} \right)_s  
+\left(\begin{array}{c}
\frac{uv}{v-u}\\ 
\frac{uv}{u-v}
\end{array} \right), \label{gt}
\end{equation}
there being no $a$-term! It is of non-homogeneous hydrodynamic systems of Tsarev-Gibbons type  \cite{gt,ff} and  it has one obvious hydrodynamic type conserved density $(u+v)$. Moreover, according to the theory of Poisson commuting Hamiltonians \cite{ff}, one can also find a conserved density of first derivatives:
\[(u-v)[(\frac{u_s}{u})^2-(\frac{v_s}{v})^2 ].\]
\indent For a given affinor $v_i^j(u)$ of a hydrodynamic system
\begin{equation}
\frac{\pa{u_i}}{\pa y}=v_i^j(u)\frac{\pa u_j}{\pa x},\label{hy}
\end{equation}
one can define the Nijenhuis tensor $N_{ij}^k(u)$ \cite{ni}
\[N_{ij}^k=v_i^s \frac{\pa v_j^k}{\pa u_s}-v_j^s \frac{\pa v_i^k}{\pa u_s}+v_s^k \frac{\pa v_i^s}{\pa u_j}-v_s^k \frac{\pa v_j^s}{\pa u_i}. \]
Then we can define the corresponding Haantjes tensor $H_{jk}^i(u)$ \cite{h}
\[H_{jk}^i=(N_{qp}^iv_{k}^q-N_{kp}^q v_{q}^i)v_j^p-(N_{qj}^p v_{k}^q-N_{kj}^q v_{q}^p)v_p^i.\]
The system \eq{hy} is diagonalized if and only if the Haantjes tensor $H_{jk}^i(u)$ vanishes identically and 
all the eigenvalues of the affinor $v_{i}^j(u)$ are real and distinct. That is, there exist $N$-functions $\la_n$(Riemann invariants), depending on the variables $u_i$, in which the equation \eq{hy} is diagonalized
\begin{equation}
\frac{\pa \la_n}{\pa y}=V_n \frac{\pa \la_n}{\pa x},\label{ri}
\end{equation}
where $V_n$ are the eigenvalues of the matrix $v_i^j(u)$, called  characteristic speed. For the hydrodynamic system \eq{c1} and \eq{c2} of the reduced dDym, the Haantjes tensor $H_{jk}^i(u)$ vanishes identically whenever the system
\eq{n1}and \eq{n2} is satisfied. Therefore, any consistent reduction of dDym is diagonalizable and then we can use Riemann
invariants to discuss problem further. \\ 
\indent Finally, one remarks that as in the case of dKP hierarchy \cite{KG}, a similar argument shows that for the reduced  dDym hierarchy \eq{dy} we also have the Kodama-Gibbons formulation: the Riemann invariants are 
\begin{equation}
\la_n=\la(q_n), \quad \mbox{where} \quad \frac{\pa \la}{\pa p}(q_n)=0, \quad n=1,2,\cdots,N \label{rie}
\end{equation}
and then the hierarchy \eq{dy} can be expressed as
\[\frac{\pa \la_n}{\pa t_m}=\hat \Om_{m}(\hat V_n)\frac{\pa \la_n}{\pa x},\]
where $\hat V_n=(q_1, q_2, \cdots, q_N)$ and $\hat \Om_{m}(\hat V_n)=\frac{d \Om_{m}(p)}{dp}\vert_{p=\hat V_n}.$  
\section{semi-Hamiltonian property}
In this section, one will prove the semi-Hamiltonian property (Tsarev's condition)\cite{t} of the reduced dDym hierarchy
\eq{dy} using Riemann invariants. Suppose that each moments $A_n$  can be expressed as the Riemann's invariants $\la_i$,
which satisfy the equation($t_2=y$)
\begin{equation}
\frac{\pa \la_i}{\pa y}=V_i \frac{\pa \la_i}{\pa x},\label{ri}
\end{equation}
where 
\[V_i=2A_{-1}^2 q_i, \quad i=1,2,3\cdots, N.\]
Then the moment equations \eq{ha} can be written as
\begin{equation}
V_i  A_{n, \la_i}=2A_{-1}^2 A_{n+1, \la_i}+(n+1) A_{n+1}(A_{-1}^2)_{\la_i}, \label{ha1}
\end{equation}
On the other hand, from \eq{dy}, we also have, equivalent to \eq{ha1},
\[\frac{\pa \la}{\pa y}=\frac{\pa \la}{\pa p}(A_{-1}^2)_{x}p^2-2\frac{\pa \la}{\pa x}A_{-1}^2p\]
and then, using \eq{ri}, one obtains, after reshuffling terms, 
\begin{equation}
\frac{\pa \la}{\pa \la_i}=p^2 \frac{\pa \la}{\pa p}(A_{-1}^2)_{\la_i} \frac{1}{V_i+2A_{-1}^2p}. \label{qu}
\end{equation}
Cross-differentiating, one has ($\phi=A_{-1}^2$)
\begin{eqnarray*}
&&\phi_{\la_i \la_j}(V_i+2\phi p)(V_j+2\phi p)^2-\phi_{\la_i}\phi_{\la_j} 2\phi p^2(V_i+2\phi p)-(V_j+2\phi p)^2 \phi_{\la_i} \frac{\pa V_i}{\pa \la_j} \\
&&-2p\phi_{\la_i}\phi_{\la_j}(V_j+2\phi p)^2 \\
&&=\phi_{\la_i \la_j}(V_j+2\phi p)(V_i+2\phi p)^2-\phi_{\la_i}\phi_{\la_j} 2\phi p^2(V_j+2\phi p)-(V_i+2\phi p)^2 \phi_{\la_j} \frac{\pa V_j}{\pa \la_i} \\
&&-2p\phi_{\la_i}\phi_{\la_j}(V_i+2\phi p)^2.
\end{eqnarray*}
Letting $p=\frac{-V_j}{2\phi}$, we obtain 
\begin{equation}
\frac{\pa V_j}{\pa \la_i}=\frac{1}{2}(\ln \phi)_{\la_i}[\frac{V_i V_j}{V_i-V_j}+V_j], \quad i \neq j .\label{one}
\end{equation}
On the other hand, comparing the coefficients of p-power and using \eq{one}, we get the only equation 
\begin{equation}
\phi_{\la_i \la_j}=\frac{\phi_{\la_i}\phi_{\la_j}}{\phi}[\frac{V_iV_j}{(V_j-V_i)^2}+1], \quad i \neq j .\label{two}
\end{equation}
The higher moments $A_n$, with $n \geq 0$ can be solved recursively using \eq{ha1}. These equations \eq{one}\eq{two}
are compatible and their solutions are parameterized by $2N$ functions of a single variable. \\
\indent A direct calculation, using MAPLE,  confirms that the reduced equation \eq{ri} is semi-Hamiltonian, that is,
\begin{eqnarray*}
&&\frac{\pa}{\pa \la_k} \left(\frac {\frac{\pa V_j}{\pa \la_i}}{V_j-V_i} \right) = \frac{\pa}{\pa \la_k} \left(\frac{1}{2}(\ln \phi)_{\la_i}[1-(\frac{V_i}{V_j-V_i})^2 ]\right) \\
&&=\frac{\pa}{\pa \la_i} \left(\frac{1}{2}(\ln \phi)_{\la_k}[1-(\frac{V_k}{V_j-V_k})^2 ]\right)=\frac{\pa}{\pa \la_i} \left(\frac {\frac{\pa V_j}{\pa \la_k}}{V_j-V_k} \right),
\end{eqnarray*}
for $i,j,k$ all distinct. Then the reduced equations \eq{ri} are thus integrable by the generalized hodograph transformation \cite{t}.

\section{Algebraic Equations and Special Reductions}
To investigate the reduction problems, we introduce $A=2\ln A_{-1}$ to put \eq{one} and \eq{two} in a more compact 
form($i \neq j$)
\begin{eqnarray*}
\frac{\pa V_j}{\pa \la_i}&=&\frac{1}{2}A_{\la_i}[\frac{V_i V_j}{V_i-V_j}+V_j]  \\
A_{\la_i \la_j}&=& A_{\la_i}A_{\la_j}\frac{V_iV_j}{(V_j-V_i)^2},
\end{eqnarray*}
or, noting that  $V_i=2A_{-1}^2 q_i$, 
\begin{eqnarray}
\frac{\pa q_j}{\pa \la_i}&=&\frac{1}{2}A_{\la_i}[\frac{q_i q_j}{q_i-q_j}-q_j]  \label{A1} \\
A_{\la_i \la_j}&=& A_{\la_i}A_{\la_j}\frac{q_iq_j}{(q_j-q_i)^2}, \label{A2}
\end{eqnarray}
Then as in \cite{gt} for the dKP case, we impose two further restrictions on the reduced system \eq{A1} and \eq{A2}. First, from the form \eq{A1}, we require that the reduced system is translation-invariant in the sense that 
\[\la_i \to \la_i+c \Rightarrow q_i \to q_i, \quad  A \to A\]
or,equivalently,
\begin{eqnarray}
\de q_i&=& 0 \label{tr1} \\
\de A &=& 0 \label{tr2}, 
\end{eqnarray}
where $\de = \sum_{i=1}^N \frac{\pa}{\pa \la_i}$. Secondly, we require the homogeneity of the functions $A$ and $q_j$ in the variables $\la_i$. $A$ should be of weight 0, and the $q_j$ of weight $-1$. Hence 
\begin{eqnarray}
R q_i&=& -\frac{1}{\ka}q_i  \label{tr3} \\
R \frac{\pa A}{\pa \la_i} &=&-\frac{\pa A}{\pa \la_i} \label{tr4}, 
\end{eqnarray}
where $\ka$ is a positive integer and $R = \sum_{i=1}^N \la_i \frac{\pa}{\pa \la_i}$. Plugging \eq{A2} into \eq{tr2}\eq{tr4} and eliminating the second derivative, we obtain
\begin{eqnarray*}
-\frac{\pa A}{\pa \la_i} &=& R \frac{\pa A}{\pa \la_i}=\sum_{j=1, j \neq i}^N \la_j \frac{\pa^2 A}{\pa \la_i \la_j}+\la_i
\frac{\pa^2 A}{\pa \la_i^2 } \\
&=& \sum_{j=1, j \neq i}^N \la_j \frac{\pa^2 A}{\pa \la_i \la_j} +\la_i (-\sum_{j=1, j \neq i}^N  \frac{\pa^2 A}{\pa \la_i \la_j})  \\ 
&=& \sum_{j=1, j \neq i}^N (\la_j-\la_i)A_{\la_j}A_{\la_i} \frac{q_iq_j}{(q_j-q_i)^2}.
\end{eqnarray*}
Hence, either $A_{\la_i}=0$ or 
\begin{equation} 
\sum_{j=1, j \neq i}^N (\la_j-\la_i)A_{\la_j}\frac{q_iq_j}{(q_j-q_i)^2} = -1. \label{B2}
\end{equation}
Similarly, plugging \eq{A1} into \eq{tr1} and \eq{tr3}, we get 
\begin{eqnarray}
-\frac{1}{\ka} q_i &=& Rq_i=\sum_{j=1, j \neq i}^N \la_j \frac{\pa q_i}{\pa \la_j}+\la_i \frac{\pa q_i}{\pa \la_i} \no \\
&=& \sum_{j=1, j \neq i}^N \la_j \frac{\pa q_i}{\pa \la_j}-\la_i\sum_{j=1, j \neq i}^N  \frac{\pa q_i}{\pa \la_j} \no \\
&=& \sum_{j=1, j \neq i}^N (\la_j-\la_i) \frac{\pa q_i}{\pa \la_j} \no \\
&=& \frac{1}{2} \sum_{j=1, j \neq i}^N (\la_j-\la_i) \frac{\pa A}{\pa \la_j}[\frac{q_i q_j}{q_j-q_i}-q_i]. \label{sc}
\end{eqnarray}
The systems \eq{B2} and \eq{sc} form a $2N$ algebraic equations for $2N$ unknowns, the $q_i$ and $\frac{\pa A}{\pa \la_i}$. The solutions of the system are essentially unique. With two Riemann invariants and $\ka =1$, a simple calculation can yield by carefully choosing the integration constants 
\begin{equation}
q_1=-q_2=\frac{4}{\la_1-\la_2},  \quad  A_{-1}=\frac{(\la_1-\la_2)^2}{16}. \label{sol}
\end{equation}
Hence from \eq{ri} we have 
\begin{equation}
 \left(\begin{array}{c}
\la_1 \\
\\
 \la_2 
\end{array} \right)_y
=\left(\begin{array}{cc}
\frac{(\la_1-\la_2)^3}{32}& 0 \\
\\
0&-\frac{(\la_1-\la_2)^3}{32}
\end{array} \right)
\left(\begin{array}{c}
\la_1 \\ 
\\
\la_2
\end{array} \right)_x.  \label{ho}
\end{equation}
This equation \eq{ho} will correspond to the reduction of Lax operator constructed by Riemann-Hilbert method \cite{ct,ct1}
\begin{equation}
\la(p)=A_{-1} p+A_0+\frac{1}{p} .\label{lax}
\end{equation}
If we consider more general Lax reductions of the form for positive integers $m$ and $\ka$ \cite{ct,ct2}
\begin{eqnarray}
\la(p)^m&=&A_{-1}^m p^m+w_{m-1}p^{m-1}+w_{m-2}p^{m-2}+ \cdots \label{form} \\
        &+&w_{-\ka+1}p^{-\ka+1}+   p^{-\ka}, \quad  m+\ka =N , \no 
\end{eqnarray}
then from \eq{rie} it's not difficult to see that \eq{form} satisfies the conditions \eq{tr1}\eq{tr2}\eq{tr3}\eq{tr4}. Hence its corresponding characteristic speeds and $A_{-1}$ will be one of the solution of the $2N$ algebraic equation \eq{B2} and \eq{sc}. It would be interesting to know whether any other solutions exist.\\
\indent Next, one  generalizes the operator $R$ in \eq{tr3} or  \eq{tr4} to the following forms 
\begin{eqnarray}
\hat R &=& \sum_{k=1}^N \hat g_k(\vec \la)\pa_{\la_k} \hat h_{k}(\vec \la) \no \\
\hat R V_i &=& -\frac{1}{\ka} V_i \label{tr5}
\end{eqnarray}
and 
\begin{eqnarray}
\tilde  R &=& \sum_{k=1}^N \tilde g_k(\vec \la)\pa_{\la_k} \tilde h_{k}(\vec \la) \no \\
\tilde  R \frac{\pa A}{\pa \la_i} &=&-\frac{\pa A}{\pa \la_i} \label{tr6},
\end{eqnarray}
where $\vec \la=(\la_1, \la_2, \cdots, \la_N)$ and $(\hat g_k, \hat h_k),(\tilde g_k, \tilde h_k)$ are arbitrary functions satisfying the conditions compatible with \eq{tr1} \eq{tr2}
\begin{equation}
\de \hat g_k=\de \hat h_k=\de \tilde g_k=\de \tilde h_k=0. \label{de}
\end{equation}
Then a similar calculation can get the following $2N$ algebraic equations generalizing \eq{B2} \eq{sc}
\begin{eqnarray}
\sum_{j=1, j\neq i}^N (\hat g_j \hat h_j-\hat g_i \hat h_i)A_{\la_j}[\frac{q_jq_i}{q_j-q_i}+q_i] = -\frac{1}{\ka} V_i(1+\ka \sum_{j=1}^N \hat g_j \frac{\pa \hat h_j}{\pa \la_j})  \label{B3}  \\
\sum_{j=1, j\neq i}^N (\tilde g_j \tilde h_j-\tilde  g_i \tilde h_i)A_{\la_j}\frac{q_iq_j}{(q_j-q_i)^2} = -(1+\sum_{j=1}^N \tilde g_j \frac{\pa \tilde h_j}{\pa \la_j}). \label{B4} 
\end{eqnarray}
For $N=2$ and $\ka=1$, simple calculations  can show that a suitable choice of $(\hat g_1,\hat h_1),(\hat g_2,\hat h_2),(\tilde  g_1,\tilde  h_1),(\tilde  g_2,\tilde  h_2)$, not unique,  can also obtain the solution \eq{sol}. \\
\indent Finally, one notices  that if we let 
\[ \hat h_j=\tilde h_j=1 \]
for $j=1,2, \cdots, N$, then we get the weaker condition 
than \eq{de}
\[\de \hat g_k=\de \tilde g_k.\]
Hence the equations \eq{B3}\eq{B4} reduce to 
\begin{eqnarray*}
\sum_{j=1, j\neq i}^N (\hat g_j -\hat g_i )A_{\la_j}[\frac{q_jq_i}{q_j-q_i}+q_i] &=& -\frac{1}{\ka} V_i   \\
\sum_{j=1, j\neq i}^N (\tilde g_j -\tilde  g_i )A_{\la_j}\frac{q_iq_j}{(q_j-q_i)^2} &=& -1
\end{eqnarray*}
If  $\hat g_j=\tilde g_j=\la_j$, then we can obtain \eq{B2} and \eq{sc}.
\section{concluding remarks}
\indent We prove the semi-Hamiltonian property of reductions for the d-Dym and find some particular solutions invariant under translation  and homogenity. In spite of the results obtained , there are some interesting issues deserving investigations.
\begin{itemize}
\item The integrability and solution structure of the equation \eq{gt} (or \eq{mo}) is unclear \cite{bm, ff, fg, fh}.
It is not difficult to see that \eq{gt} can be extended to :
\[\frac{\pa u_i}{\pa r}=r_i \frac{\pa u_i}{\pa s}+ \frac{u_1 u_2 \cdots u_N}{\Pi_{k \neq i}(u_k-u_i)}, \mbox{where} \quad 
r_i= (\sum_{k=1}^N u_k)- u_i. \]
One hopes  issue these problems elsewhere.
\item In \cite{kr2}, the algebraic reductions for dKP are found and in \cite{bk,ly}, the waterbag reduction(non-algebraic) for dKP  is also found.  If we define 
\[A_n=\int_{-\infty}^{\infty} q^{n}f(q,x,y) dq ,\quad n=-1,0,1,2 ,\cdots \]
then we obtain
\begin{eqnarray}
\la=p^2  (P\int_{-\infty}^{\infty} \frac{f/q}{p-q} dq),  \quad f=f(q,x,y), \label{va}
\end{eqnarray}
where $P \int$ denotes the Cauchy principal value of the integral. Also,from \eq{va}, the "distribution"  function $f(x,y,q)$ must satisfy the Vlasov-like equation \cite{g,pb, Z}
\begin{equation}
f_y=2A_{-1}^2 q f_x-(A_{-1}^2)_x q^2 f_q=\{f, A_{-1}^2 q^2 \}_{x,q}.\label{la1}
\end{equation}
Comparing \eq{la1} and \eq{la}, we can assume $f=F(\la)$ for any function $F$. Hence the Lax operator $\la$ will satisfy the non-linear singular integral equation
\[\la=p^2 (P\int_{-\infty}^{\infty} \frac{F(\la)/q}{p-q} dq). \]
This equation will help us find the (non-)algebraic reductions \cite {gt1, ma}, study the initial value problem for d-Dym as in the case of dKP \cite{yg,ly} and it needs further investigations.
\end{itemize}
{\bf Acknowledgments\/} \\
The author is grateful for M.Pavlov's  and M.H. Tu's  stimulating discussions on hydrodynamic system and its reductions.

\end{document}